\documentclass[prb,preprint]{revtex4-1} 

\usepackage{amsmath}  
\usepackage{amsfonts} 
\usepackage{graphicx} 

\begin{document}


\title{Solutions of the Schroedinger equation for piecewise  harmonic potentials: remarks on the asymptotic behavior of the wave functions}


\author{Francisco D. Mazzitelli, Mart\'in D. Mazzitelli and Pedro I. Soubelet}
\affiliation{Centro At\'omico Bariloche and Instituto Balseiro, 
Comisi\'on Nacional de Energ\'\i a At\'omica, 8400 Bariloche, Argentina.}

\date{today}

\begin{abstract} 
We discuss the solutions of the Schroedinger equation for  piecewise potentials,  given by the harmonic oscillator potential for
$\vert x\vert >a$ and an arbitrary function for $\vert x\vert <a$, using elementary methods. 
The study of this problem sheds light on usual
errors  when discussing the asymptotic behavior of the eigenfunctions of the quantum harmonic oscillator and can also be used for the analysis 
of the eigenfunctions of the hydrogen atom. 
We present explicit results for the energy levels of a potential of this class, used to model the confinement of electrons in nanostructures. 
\end{abstract}

\maketitle 

\section{Introduction}\label{sec:intro}

The Schroedinger equation for the linear harmonic oscillator reads
\begin{equation}
\frac{d^2\psi}{dz^2}+({\mathcal E}-\frac{z^2}{4})\psi=0\, ,
\label{Schroed}
\end{equation}
where $z=\sqrt{2 m \omega/\hbar}\, x$, $\mathcal{E}=E/(\hbar\omega)$, $x$ is the coordinate of the oscillator, and $m$ and $\omega$ its mass and frequency, respectively.
The traditional approach for solving this equation consists in proposing a solution of the form
\begin{eqnarray}
\psi(z)&=&e^{-z^2/4}\left[\sum_{n\geq 0} a_{2n}z^{2n}+  \sum_{n\geq 0} a_{2n+1}z^{2n+1}\right]
\nonumber\\
&\equiv& e^{-z^2/4}\left[a_0 \, S_{even}(z)+a_1 \, S_{odd}(z)\right]\, ,
\label{prop}
\end{eqnarray}
and obtain recurrence relations for the coefficients
\begin{equation}
a_{n+2}=\frac{n-{\mathcal E}+1/2}{(n+1)(n+2)}a_n\, .
\label{rec}
\end{equation}
Note that these are two sets of independent recurrence relations for the even and odd coefficients, which are fully determined once one fixes $a_0$ and $a_1$.

The energy eigenvalues are obtained by imposing the condition that  $\vert\psi(z)\vert^2$ should be integrable. In many textbooks 
\cite{text1,text2,text3,text4,text5,text6,text7,text8} 
it is remarked 
that, for large $n$,  $a_{2n+2}/a_{2n}\simeq a_{2n+3}/a_{2n+1}\simeq 1/(2n)$,  which is also the behavior for the coefficients of the series of  $e^{z^2/2}$. Then, it is
(incorrectly) argued \cite{text1,text2,text3,text4} that, for large $z$, 
\begin{equation}
S_{even}\simeq S_{odd}/z \simeq e^{z^2/2}\, ,
\label{incor}
\end{equation} 
and therefore $\vert\psi(z)\vert^2$ would not be integrable, unless 
the series include only a finite number of terms. As a consequence,  the allowed energy eigenvalues are ${\mathcal E}=n+1/2$, for some nonnegative integer $n$.

Long time ago it has been pointed out that, although the conclusion for the eigenvalues  is correct, the argument is wrong. \cite{Buchdahl} It is not true that, if two power series with coefficients
$a_n$ and $b_n$ are such that $a_{n+1}/a_n\simeq b_{n+1}/b_n$ for large $n$, then  they have the same behavior for large arguments. This is 
implicitly assumed in other textbooks,  \cite{text5,text6,text7,text8} where it is pointed out that $a_{2n+2}/a_{2n}\simeq 1/(2n)$ is the behavior of the coefficients of the series
of $z^k e^{z^2/2}$ for {\it any value of} $k$. This property is used to argue that  the asymptotic behavior of the odd and even series
 should be of this form for some particular values of $k$. Note that this claim is at least
incomplete, since $z^k e^{z^2/2}$ admits a representation in powers of $z$ (for all  $z$),  only if $k$ is a natural number (or eventually an integer, if one considers
also Laurent series). Moreover, in principle there could exist other functions with different asymptotic behavior and the same  ratio
$a_{2n+2}/a_{2n}$ in the large $n$ limit.

A correct reasoning is as follows. \cite{Bowen} If $a_{n+1}/a_n > b_{n+1}/b_n$ and $a_n, b_n >0$ for $n\geq N$, then one can show that
\begin{equation} \label{bound}
\sum_{n\geq 0} a_{n}z^{n}\geq k \sum_{n\geq 0} b_{n}z^{n} + P(z)\, ,
\end{equation}
where $k$ is a positive constant and $P(z)$ a polynomial of degree $N$. One can use this bound to show that the odd and even series
in Eq.~\eqref{prop} diverge faster than $e^{\alpha z^2}$ with $\alpha<1/2$, unless they contain a finite number of terms. From this property one can derive the allowed eigenvalues for the harmonic oscillator. 

In the present paper we discuss a related problem: the Schroedinger equation in the presence of piecewise potentials that coincide with the harmonic oscillator potential for $\vert x  \vert >a$. The analysis of this potential makes more evident the usual mistakes in the discussions of the asymptotic behavior of the wave functions, as the following (erroneous)  argument shows:  if  ${\mathcal E}-1/2$ were not an integer, as  the odd and even series cannot cancel each other for $z\to +\infty$ (Eq.~\eqref{incor}),  the wave function would not be quadratically integrable. Therefore, ${\mathcal E}-1/2$ must be a nonnegative integer, and the eigenvalues for the piecewise potentials would coincide with those of the harmonic oscillator, irrespective of the form of the potential for  $\vert x  \vert <a$. This is obviously nonsense. The error in the argument goes back to the assumed asymptotic behavior in Eq.~\eqref{incor}. As we will see,   $S_{even}/S_{odd}\to const$ as $z\to + \infty$, for any value of  ${\mathcal E}$ such that ${\mathcal E}-1/2\neq 0,1,2,... \, .$

It is worth to note that, for solving  this problem,  it is not enough to obtain a lower bound for the series:  the leading behavior of both series is needed. As this behavior is not difficult to obtain,  it will be useful even when discussing the usual  quantum harmonic oscillator.  Moreover, when considering the radial Schroedinger equation for the hydrogen atom, one also encounters vague arguments in the analysis of the asymptotic behavior of the solutions. Our results  shed light  on the discussion on this and related problems.

Piecewise potentials involving the harmonic potential have been considered before by other authors. \cite{similar1, similar2, similar3,similar4,similar5} In some works,  
\cite {similar1,similar2,similar3} 
 the harmonic part of the potential is restricted to a bounded region ($\vert x\vert < a$ in our notation), the opposite situation of the one considered here, and therefore the discussion of the asymptotic behavior of the series  Eq.~\eqref{prop}
is not relevant there. In other works,  \cite{similar4, similar5} the authors consider the combination of an harmonic potential for $x>a$ and a finite potential step for $x<a$. In this case, 
the analysis of the large-$x$ behavior of the solutions is relevant, and could be discussed  using the elementary methods proposed below. Alternatively, in Ref.\  \onlinecite{similar4} the problem is tackled  solving Schroedinger equation in terms of special functions, while in Ref.\ \onlinecite {similar5} the eigenvalue equation is solved using an integral representation method.

\section{Schroedinger equation with piecewise harmonic potentials}  

Let us now consider the Schroedinger equation
\begin{equation}
\frac{d^2\psi}{dz^2}+({\mathcal E}-V(z))\psi=0\, ,
\label{SchroedV}
\end{equation}
with
\begin{equation}
\label{ppotential}
V(z)=\left\{\begin{matrix}
\frac{1}{4} (z+l)^2\,\,\,\,\, &&z<-l, 
\\ f(z) \,\, && -l<z<l ,
\\ \frac{1}{4} (z-l)^2 \,\, &&z>l,
\end{matrix}\right.
\end{equation}
where $f(z)$ is an arbitrary function and $l=\sqrt{2 m \omega/\hbar}\, a$. The potential is harmonic for $\vert z\vert >l$ and arbitrary otherwise.

Let us first analyze  the asymptotic behavior of the solutions of Eq.~\eqref{SchroedV}.  As the potential is defined in three different regions,  one can
study the behavior for $z<-l$ and $z>l$ separately. It will be enough to analyze the asymptotic behavior in the region  $z > l$. We
 introduce the notation $y=z-l$. Given the form of the potential, one expects that, for $y\to +\infty$:
\begin{equation}\label{asymp}
\psi(y)\simeq y^\gamma e^{\beta y^2}\, ,
\end{equation}
for some constants $\beta$ and $\gamma$. Indeed, inserting this ansatz into Eq.~\eqref{SchroedV} we obtain
\begin{equation}\label{betagamma}
4 \beta^2-\frac{1}{4} + y^{-2} ({\mathcal E}+2\beta(1+2\gamma)) +O(y^{-4})=0\, ,
\end{equation}
that is satisfied, in the limit
$y\gg 1$,  when
\begin{equation}
\beta^2= \frac{1}{16}\quad  \gamma=-\frac{1}{4\beta} ({\mathcal E}+ 2 \beta)\, .
\label{asympcoef}
\end{equation}
We conclude that the Schroedinger equation has a solution that converges at $y\to+\infty$ ($\beta=-1/4, \gamma={\mathcal E}- 1/2$) and a linearly independent solution that diverges in the same
limit ($\beta=1/4, \gamma=-{\mathcal E}-1/2$). The analysis could be pursued systematically by assuming
 \begin{equation}\label{asymp2}
\psi(y)\simeq y^\gamma e^{\beta y^2}(1+\frac{\gamma_1}{y}+\frac{\gamma_2}{y^2}+...)\, ,
\end{equation}
but this will not be necessary for what follows.

In the usual discussions of the asymptotic behavior  of the solutions of the harmonic oscillator, only the leading term is kept in Eq.~\eqref{betagamma}. This  gives $\beta=\pm 1/4$, and no information on $\gamma$. 

We now propose a solution of  Eq.~\eqref{SchroedV} for $y>0\,  (z > l)$  of the form given in Eq.~\eqref{prop}, with $y$ instead of $z$. As we expect 
that the eigenvalues for the piecewise potentials will differ from those of the usual harmonic oscillator, in what follows we will assume that  
${\mathcal E}-1/2\neq 0,1,2,...\,  .$  The usual eigenvalues will be obtained in the limiting case $l\to 0$.

The bounds Eq.~\eqref{bound} on the odd and even series
imply that both
$e^{-y^2/4}S_{even}(y)$ and  $e^{-y^2/4}S_{odd}(y)$ diverge as $y\to +\infty$.   However, the existence of solutions with the asymptotic behavior given
in Eq.~\eqref{asymp} with $\beta=-1/4$ implies that there should be a {\it unique choice} of $a_1/a_0$ such that the combination
\begin{equation}\label{psiy}
\psi(y)=e^{-y^2/4}\left[ a_0  S_{even}(y)+ a_1  S_{odd}(y)\right]
\end{equation}
converges as $y\to\ +\infty$. When $a_1/a_0\equiv a_*$ is  properly chosen,
 the linear combination of the two divergent series becomes convergent.  It is important to remark that this should happen for any value of ${\mathcal{E}}$. The value of $a_*$ is clearly unique, otherwise one would obtain two convergent,  linearly independent solutions of the differential equation, and 
 the divergent solutions would not exist.

In the next section we will obtain the precise value of $a_*$. Assuming that this value is known, it is easy to  find the set of equations that determines
the energy eigenvalues. We introduce the notation
\begin{equation}\label{weber}
D_{\mathcal E-1/2}(y) = S_{even}(y)+ a_* S_{odd}(y)\, .
 \end{equation}
In terms of this function, the quadratically integrable solution of Eq.~\eqref{SchroedV} can be written as
\begin{equation}
\psi(z)=\left\{\begin{matrix}
A e^{-(z+l)^2/4}D_{\mathcal E-1/2}(-(z+l))  \,\,\,\,\, &&z<-l,
\\ B \psi_1(z)+C\psi_2(z) \,\, && -l<z<l, 
\\ F e^{-(z-l)^2/4}D_{\mathcal E-1/2}(z-l)\,\, &&z>l,
\end{matrix}\right.
\label{soln}
\end{equation}
where $\psi_1$ and $\psi_2$ are two linearly independent solutions in the region $\vert z\vert <l$, while $A,B,C,$ and $F$ are constants.
The wavefunction $\psi$ and its first derivative should be continuous both at $z=\pm l$. These four conditions and the normalization of the wavefunction
determine the four constants and the allowed values of the energy.

\subsection{Calculation of $a_*$}

From the recurrence relation Eq.~\eqref{rec} one can see that
\begin{eqnarray}
a_2&=&\frac{a_0(-\mathcal E+1/2)}{2}=\frac{a_0(-\mathcal E /2+1/4)}{2\times 1/2},\nonumber\\
a_4&=&\frac{a_0(-\mathcal E+1/2)\times (-\mathcal E+1/2+2)}{2\times 3\times 4}=\frac{a_0 (-\mathcal E/2+1/4)\times (-\mathcal E/2+1/4+1)}{2^2 \times 1/2\times  3/2  \times 2 },\nonumber\\
a_6&=&=\frac{a_0 (-\mathcal E/2+1/4)\times (-\mathcal E/2+1/4+1)\times (-\mathcal E/2+1/4+2)  }{2^3 \times 1/2\times  3/2\times 5/2  \times 2\times 3 },
\end{eqnarray}
and, in general,
\begin{eqnarray}
a_{2n}&=&\frac{a_0}{2^n n!}\frac{(-\mathcal E /2+1/4)\times\cdots \times(-\mathcal E /2+1/4+ n-1)}{1/2\times3/2\times\cdots\times (1/2+n-1)}
\nonumber\\ &=&
 \frac{a_0}{2^n n!}\frac{\Gamma[1/2]}{\Gamma[-{\mathcal E}/2+1/4]}\frac{\Gamma[-{\mathcal E}/2+1/4+n]}{\Gamma[1/2+n]} \, ,\label{a_npar}
 \end{eqnarray}
 where $\Gamma[z]$ denotes the Gamma function. Note that in the last equality we have made repeated use of the well known identity $\Gamma[z+1]=z\Gamma[z]$.  Following similar steps, one can verify that
 \begin{equation}
a_{2n+1} = \frac{a_1}{2^n n!}\frac{\Gamma[3/2]}{\Gamma[-{\mathcal E}/2+3/4]}\frac{\Gamma[-{\mathcal E}/2+3/4+n]}{\Gamma[3/2+n]} \, .
\label{coef2}
\end{equation}

The large-$n$ behavior of the coefficients $a_n$ can be analyzed using
Stirling's approximation for the Gamma function at large arguments \cite{Abra}
\begin{equation} \label{Stirling}
\Gamma[z+1] \simeq z^ze^{-z}\sqrt{2\pi z}\, ,
\end{equation}
from which we obtain
\begin{equation}
\frac{\Gamma[n+b]}{\Gamma[n+c]}\simeq n^{b-c}\, .
\end{equation}
Inserting this approximation into Eq.~\eqref{a_npar} and  Eq.~\eqref{coef2} we obtain, for large $n$,
\begin{eqnarray}
a_{2n} &\simeq& \frac{a_0}{2^n n!}\frac{\Gamma[1/2]}{\Gamma[-{\mathcal E}/2+1/4]}n^{-{\mathcal E}/2-1/4}, \label{coef1approx}\\
a_{2n+1} &\simeq&  \frac{a_1}{2^n n!}\frac{\Gamma[3/2]}{\Gamma[-{\mathcal E}/2+3/4]} n^{-{\mathcal E}/2-3/4}    \, .
\label{coef2approx}
\end{eqnarray}

From Eq.~\eqref{coef1approx} and Eq.~\eqref{coef2approx} we see that the asymptotic behavior of $S_{even}$ and $S_{odd}$ 
can be studied by considering the series
\begin{equation}\label{serieS}
S(\omega)=\sum_{n=1}^\infty \frac{n^{-r}\omega^n}{n!}\, .
\end{equation}
Indeed,  if two power series with positive coefficients $A_n$ and $B_n$ are such that $A_n/B_n\to 1$ for $n\to\infty$, then they have the same asymptotic behavior. Hence, in virtue of Eq.~\eqref{coef1approx}, putting $\omega=y^2/2$, $r={\mathcal E}/2+1/4$ and multiplying by $a_0\, \Gamma[1/2]/\Gamma[-{\mathcal E}/2+1/4]$  on both sides of Eq.~\eqref{serieS}, we see that 
\begin{equation}\label{SevenS}
S_{even}(y)\simeq \frac{\Gamma[1/2]}{\Gamma[-{\mathcal E}/2+1/4]} S\left( \frac{y^2}{2} \right)
\end{equation}
 for large values of $y$. If, instead, we put  $r={\mathcal E}/2+3/4$ and  multiply by ${a_1\, \Gamma[3/2]/\Gamma[-{\mathcal E}/2+3/4]}$ on both sides of Eq.~\eqref{serieS}, we obtain 
 \begin{equation}
 \label{SoddS}
 S_{odd}(y) \simeq \frac{y\, \Gamma[3/2]}{\Gamma[-{\mathcal E}/2+3/4]}  S\left(\frac{y^2}{2}\right) 
 \end{equation}
 for large $y$.

In order to study the asymptotic behavior of $S(\omega)$, the key observation is that,
for a fixed large value of $\omega$,  
the  coefficients 
\begin{equation}
\label{coefS2}
c_n(\omega)=\frac{\omega^n}{n!}\, ,
\end{equation}
have, as a function of $n$,  a  peak at $n=\omega$. Moreover,  width of the peak is  much smaller than $\omega$. It is an interesting exercise
to verify these properties by plotting $c_n(\omega)$ as a function of $n$ for large values of $\omega$. We can prove them analytically 
using Stirling's approximation
Eq.~\eqref{Stirling} for the factorial $n!=\Gamma[n+1]$, 
and evaluating  for $n\simeq \omega$. We obtain 
\begin{equation}
n!\simeq\sqrt{2\pi\omega}e^{-n}\omega^n e^{n\ln(1+\frac{(n-\omega)}{\omega})}\, .
\end{equation}
Expanding the logarithm in the exponential in powers of $(n-\omega)/\omega$ we get
\begin{equation}
\label{coefSapprox}
c_n(\omega)\simeq \frac{e^\omega}{\sqrt{2\pi\omega}}e^{-\frac {(n-\omega)^2}{2\omega}} \, .
\end{equation}
Therefore,  for large (fixed) $\omega$, $c_n(\omega)$ is a Gaussian function of $n$, with a peak at $n=\omega$ and width $\sqrt\omega$.  
  Thus, for the relevant values of $n$ we can approximate 
$n^{-r}$ by $\omega^{-r}$ in $S(\omega)$ obtaining, for large $\omega$,  
 \begin{equation}\label{serieSSapprox}
S(\omega)\simeq \omega^{-r}\sum_{n=1}^\infty \frac{w^n}{n!}\simeq \omega^{-r}e^\omega\, .
\end{equation}  
We present a  more rigorous proof of this asymptotic behavior in the Appendix.

Taking into account Eq.~\eqref{SevenS}, Eq.~\eqref{SoddS} and Eq.\eqref{serieSSapprox} and we obtain, for large $y$, 
\begin{eqnarray}
\label{sevens} S_{even}(y) &\simeq& \frac{\Gamma[1/2]}{\Gamma[-{\mathcal E}/2+1/4]}\left[\frac{y^2}{2}\right]^{-\frac{\mathcal E}{2}-\frac{1}{4}}e^{\frac{y^2}{2}}, \\ 
\label{sodds} S_{odd}(y) &\simeq& \frac{\sqrt 2 \Gamma[3/2]}{\Gamma[-{\mathcal E}/2+3/4]}\left[\frac{y^2}{2}\right]^{-\frac{\mathcal E}{2}-\frac{1}{4}}e^{\frac{y^2}{2}}\, .
\end{eqnarray}
This calculation reproduces the asymptotic behavior of the solutions anticipated in Eq.~\eqref{asymp} and Eq.\eqref{asympcoef}. Both series lead to linearly independent solutions to the Schroedinger equation that have the same asymptotic behavior with $\beta=+1/4$ and  $\gamma=-{\mathcal E}-1/2$ (see Eq.~\eqref{asymp} and Eq.\eqref{asympcoef}). 

These two linearly independent solutions of the Schroedinger equation are not quadratically integrable, and therefore physically unacceptable. However, we know that 
a linear combination of them should produce a solution with the adequate behavior ($\beta=-1/4$). A necessary condition for this to happen is that 
the exponentially growing behavior of both series should cancel each other. Therefore, using Eq.~\eqref{psiy}, Eq.~\eqref{sevens} and Eq.~\eqref{sodds} we obtain
\begin{equation}\label{astar}
a_*=\frac{a_1}{a_0}=- \sqrt 2 \frac{\Gamma[-{\mathcal E}/2+3/4]} {\Gamma[-{\mathcal E}/2+1/4]}\, .
\end{equation}
With this result we can construct the function $D_{\mathcal E-1/2}(y)$ in Eq.~\eqref{weber}, and obtain the formal solution of the Schroedinger equation Eq.~\eqref{soln}.

We point out, for the advanced reader,  that for this particular value of $a_*$ Eq.~\eqref{psiy} reproduces the series expansion of the parabolic Weber function $D_\sigma(y)$, \cite{similar4, watson} which is the unique solution of Eq.~\eqref{SchroedV} that tends to zero as $y\to+\infty$. 

\section{Example: a particle in a ``bathtub" potential}

In order to illustrate the usefulness of the previous results, we will consider the particular case of a particle in a box bounded by harmonic walls, i.e., we will analyze the Schroedinger equation with the potential given in Eq.~\eqref{ppotential} with $f(z)=0$. These so called ``bathtub" potentials have been used as confining potentials for electrons in nanostructures, in particular when analyzing the quantum Hall effect. \cite{nano1,nano2,nano3,nano4,nano5, nano6}

As the potential is an even function, it is convenient to take as independent solutions in the region 
$ -l<z<l$ 
\begin{eqnarray}
\psi_1(z)&=& \cos\sqrt\mathcal E z,\nonumber\\
\psi_2(z)&=& \sin\sqrt\mathcal E z\, ,
\end{eqnarray}
and look for solutions of the Schroedinger equation which are either even or odd. For the even solutions we take $C=0$ in Eq.~\eqref{soln} and impose continuity of the function
and its first derivative at $z=l$. The transcendental equation that determines the energy eigenvalues is
\begin{equation}
\label{eveneigen}
\sqrt\mathcal E\tan \sqrt\mathcal E l =-\frac{D'_{\mathcal E-1/2} (0)}{D_{\mathcal E-1/2} (0)}=-a_*\, .
\end{equation}
Similarly,  for the odd solutions ($B=0$) the condition reads
\begin{equation}
\label{oddeigen}
\sqrt\mathcal E\cot \sqrt\mathcal E l =\frac{D'_{\mathcal E-1/2} (0)}{D_{\mathcal E-1/2} (0)}=a_*\, .
\end{equation}

\begin{figure}[!ht]
\includegraphics[scale=.4]{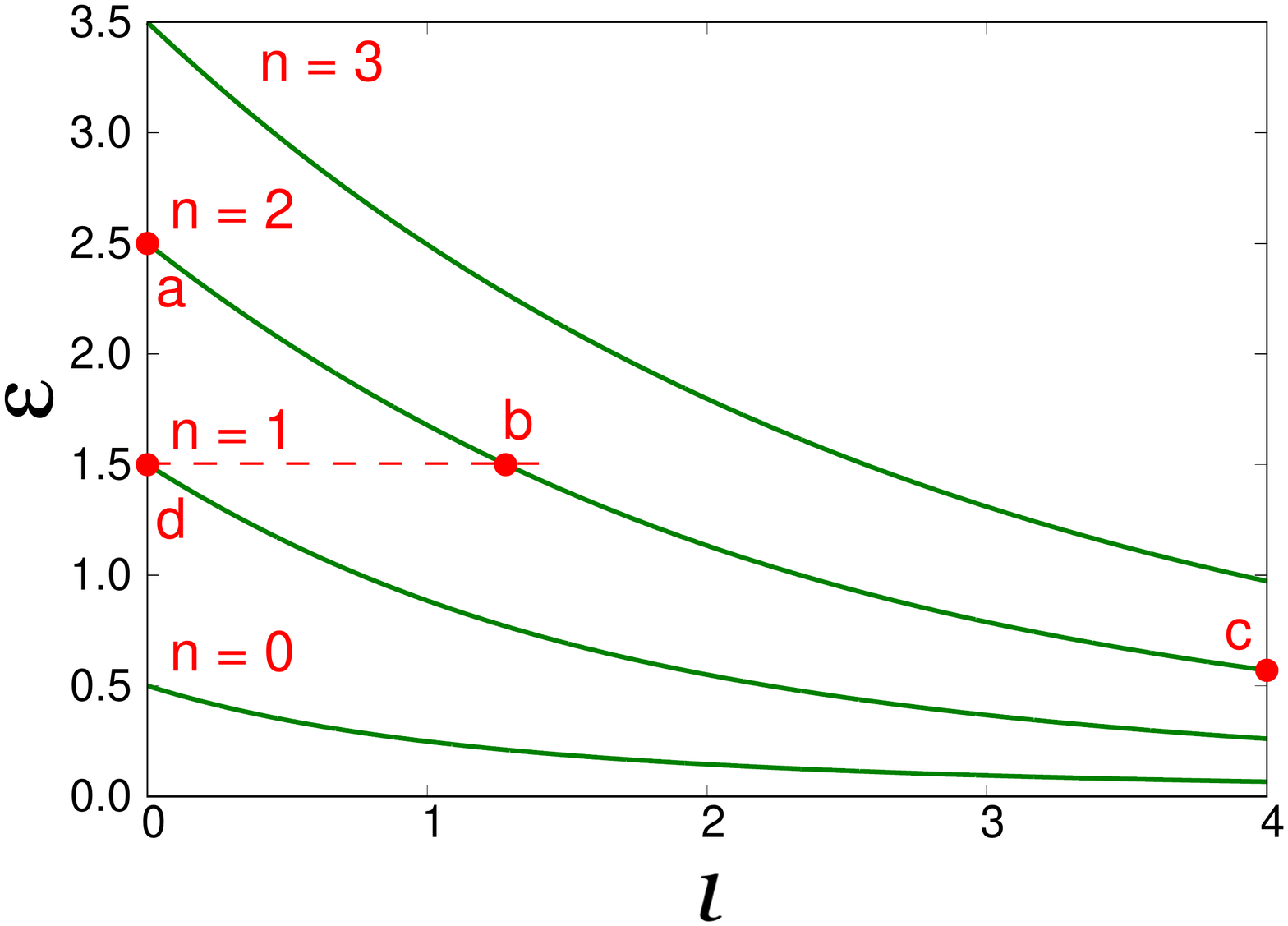} 
\caption{Eigenvalues for the ``bathtub" potential, as a function of $l$. According to the uncertainty principle, the eigenvalues are decreasing functions of 
$l$. The wave functions associated to the particular values (a), (b),(c) and (d) are plotted in Figs. 2 and 4.}
\end{figure}

In the limit $l\to 0$ one recovers the eigenvalues of the harmonic oscillator. On the one hand, for the even solutions, in this limit the condition Eq.~\eqref{eveneigen}
 reads $a_*=0$. As the Gamma function does not have zeros  for real arguments, and has poles on the non-positive integers, from Eq.~\eqref{astar} we see
 that  the argument of the Gamma function in the denominator must be a non-positive integer $-n$, and therefore ${\mathcal E}= 2 n + 1/2$, the usual eigenvalues 
for even eigenfunctions.  On the other hand,
for the odd solutions the condition  Eq.~\eqref{oddeigen} is $a_*=\infty$, which is satisfied for ${\mathcal E}= 2 n + 1+ 1/2$, i.e. the usual energy levels for odd wave functions. 

In Fig. 1 we plot the eigenvalues of the energy $\mathcal E$ as a function of  $l$. The eigenvalues start at the harmonic oscillator values $n+1/2$ for $l=0$,  and are decreasing
functions of $l$, as suggested by Heisenberg uncertainty principle. In Fig. 2 we plot the wave function of the second excited state for increasing values of $l$. At $l=0$ the wave function is the usual solution for the harmonic oscillator with energy ${\mathcal E}= 5/2$. The wave function has two nodes for all values of $l$. They are located in the harmonic region for $0<l<1.28$, and in the flat region for $l>1.28$. For this critical value of $l$, the eigenvalue of the second excited state equals $\mathcal E=3/2$, i.e. the value
of the first excited state of the usual harmonic oscillator (point (b) in Fig. 1).

\begin{figure}[!ht]
\includegraphics[scale=.4]{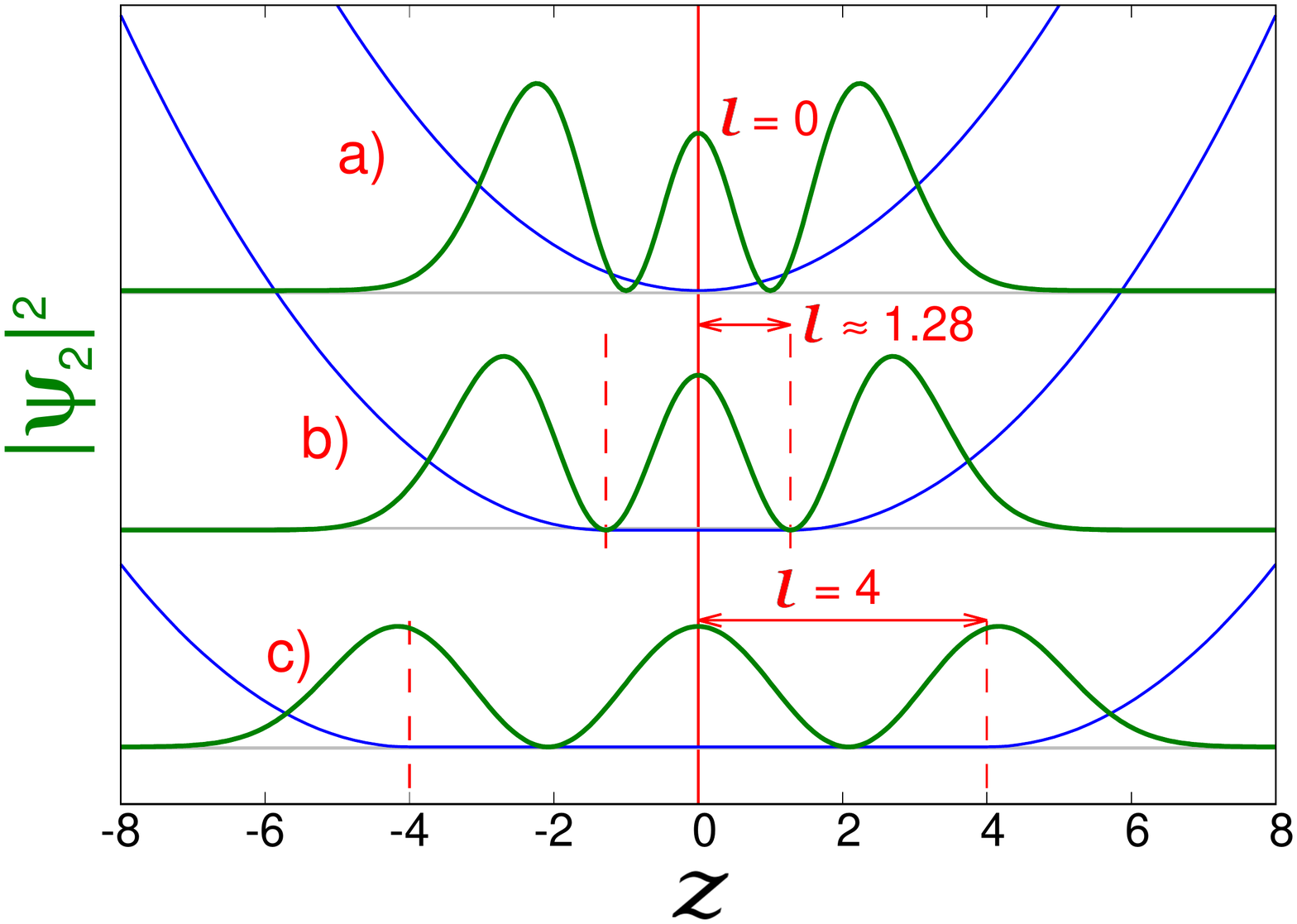} 
\caption{Plot the normalized wave function for the second excited state, for different values of $l$.  As expected, the wave function has two nodes. Note that they are located in the harmonic region for $0<l<1.28$, at $z=l$ for l=1.28 and in the flat bottom for $ l>1.28 $. The corresponding eigenvalues are given by  the points  (a), (b), and (c) in Fig. 1 }
\end{figure}

\begin{figure}[!ht]
\includegraphics[scale=.4]{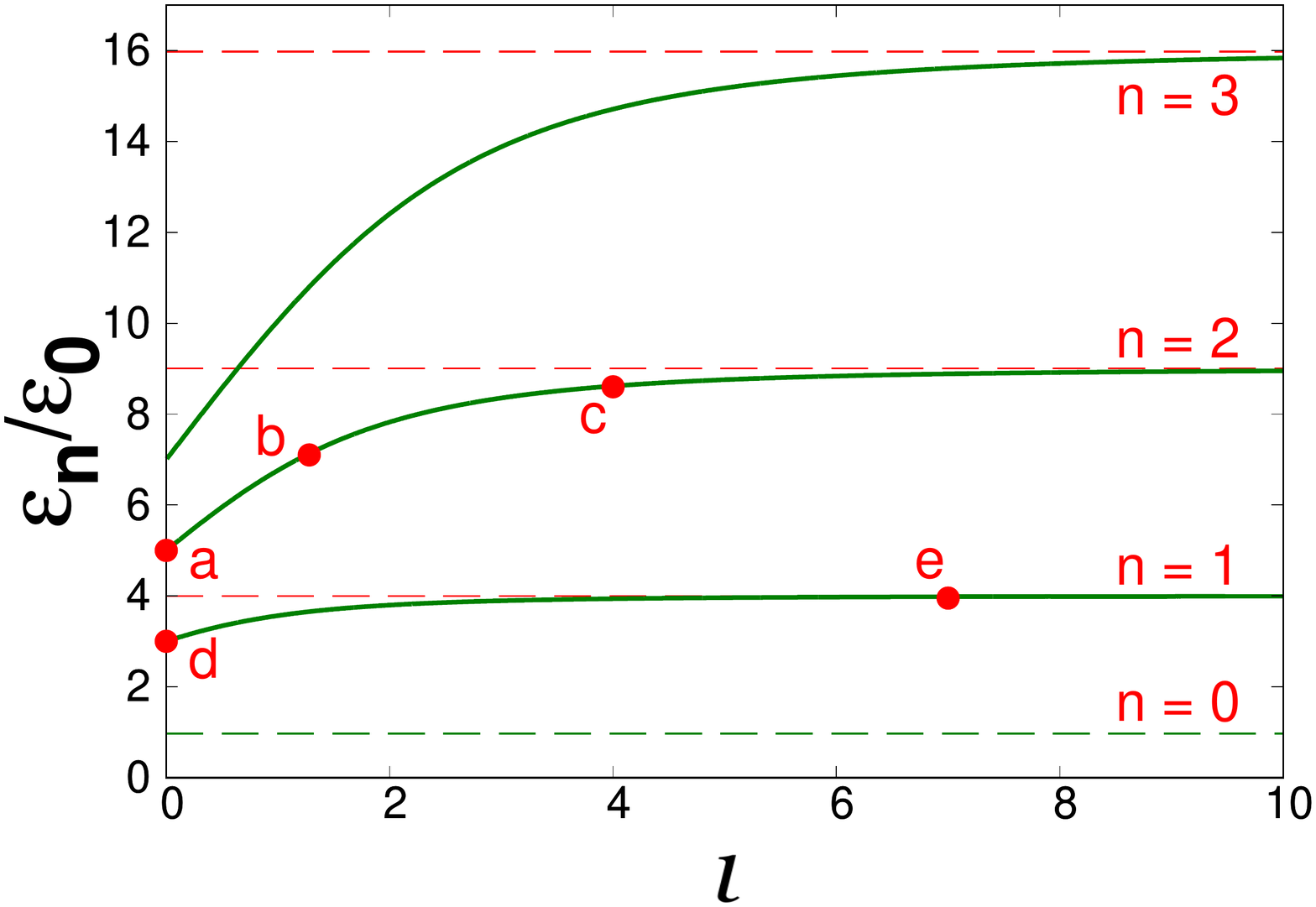} 
\caption{Eigenvalues for the ``bathtub" potential, normalized to the ground state, as a function of $l$. At large values of $l$ the spectrum
coincides with that of an infinite square well. The wave functions associated to the particular values (a), (b), (c), (d), and (e) are plotted in Figs. 2 and 4.  }
\end{figure}

An interesting property of the eigenvalues is their behavior in the limit $l\gg 1$. When $a\gg\sqrt{\hbar/2m\omega}$,
the scale of variation of the harmonic potential is much shorter than the size of the flat bottom of the potential. Therefore,  the harmonic walls act as infinite potential barriers, and  we expect the spectrum of a particle in a box, that is,  ${\mathcal E}_n/{\mathcal E}_0\simeq (n+1)^2$. This behavior is illustrated in Fig. 3.

We also expect the wave functions to evolve from those of the usual harmonic oscillator at $l=0$ to those of the infinite square well for  $l\gg 1$. This fact is illustrated in Fig. 4, where we plotted  the  first excited state for different values of $l$. Note that for large values of $l$ the wave function tends  to zero in the harmonic region,  in a spatial scale much shorter than the size of the flat bottom. 

\begin{figure}[!ht]
\includegraphics[scale=.4]{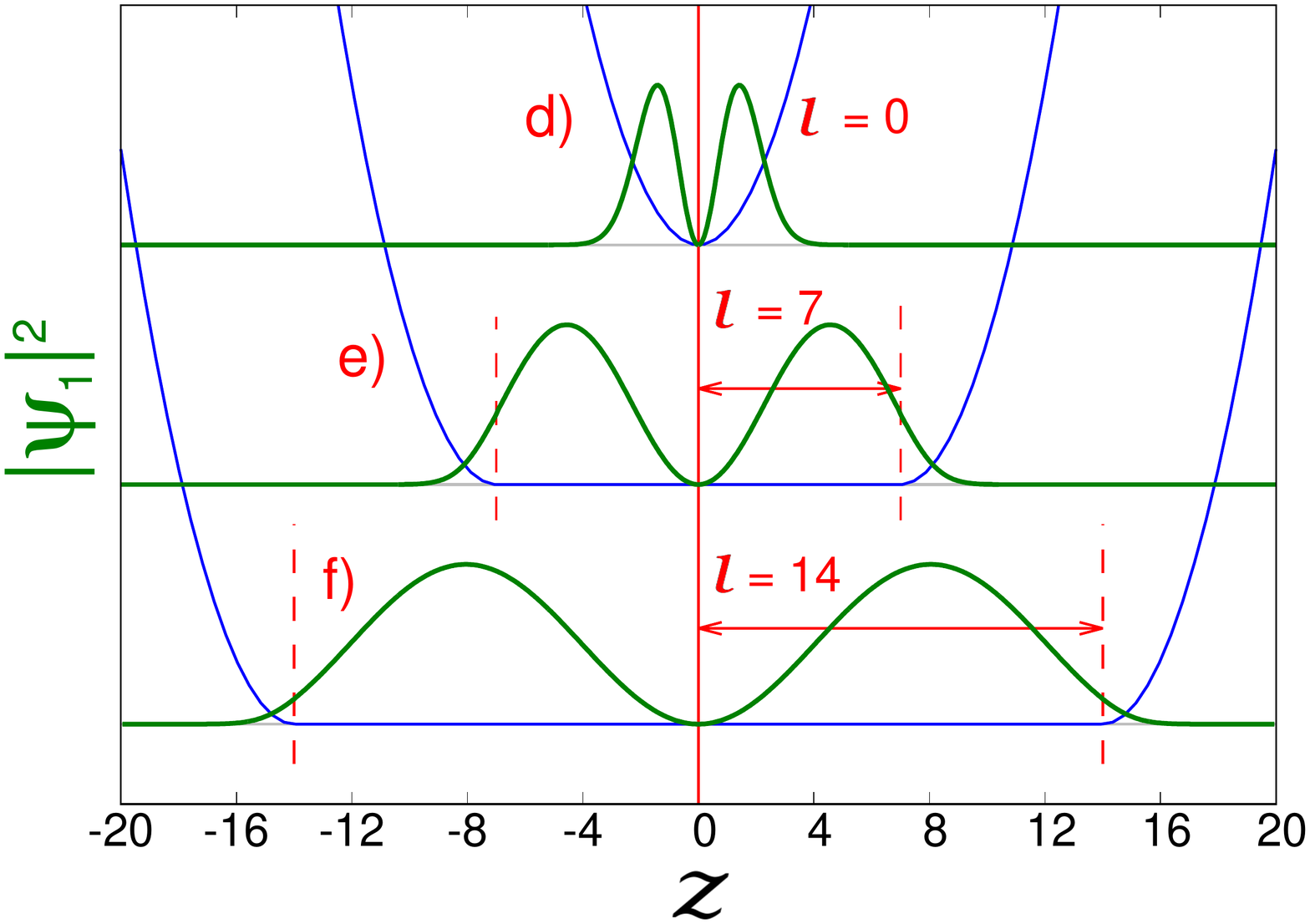} 
\caption{Plot of the wave function for the first excited state, for different values of $l$. This wave function has only one node at $z=0$ for all values
of $l$. The figure illustrates the fact that the wave function for the piecewise potential tends to that of a particle in a box, and therefore vanishes in the harmonic region in the large
$l$ limit. The corresponding eigenvalues are given by the points (d) and (e) in Fig. 3. Point (f) is out of scale in Fig. 3,  being at the right of point (e), on the curve $n=1$.  For the sake of clarity, in this figure we normalized each wave function to its maximum value.}
\end{figure}

\section{The hydrogen atom}

The remarks about the behavior of the series for the harmonic oscillator also apply to the solutions of the Schroedinger equation for the hydrogen atom.
The radial wave function is usually written as
\begin{equation}\label{radial}
\psi(\rho)=\rho^{L+1}e^{-\rho}F(\rho)\, ,
\end{equation}
where $\rho$ is a dimensionless radius, $L$ is the angular momentum,  and 
\begin{equation}
\label{F}
F(\rho)=\sum_{n\geq 0} c_n\rho^n\, .
\end{equation}

The coefficients of the power series satisfy the recurrence relation
\begin{equation}
c_{n+1}=\frac{(-\xi+ 2 L +2 + 2 n)}{(n+1)(2L+2+n)} c_n\, ,
\end{equation}
where $\xi$ is the inverse of the (dimensionless) energy. Note that, once again, for large $n$ we have $c_{n+1}/c_n\simeq 2/n$, and
one would be tempted to conclude that,  if the series does not have a finite number of terms,  $F(\rho)\simeq e^{2\rho}$ as $\rho\to\infty$. However, a 
more careful analysis along the lines of the previous sections shows that this is not the case. Indeed, the coefficients are given by
\begin{equation}
c_n=c_0\frac{2^n}{n!}\frac{\Gamma(2L+2)}{\Gamma(-\xi/2+L+1)}\frac{\Gamma(-\xi/2+L+n+1)}{\Gamma(2L+n+2)}\, ,
\end{equation}
and tend to
\begin{equation}
c_n\simeq c_0\frac{ 2^n}{n!} n^{-\xi/2-L-1}\, 
\end{equation}
for large $n$. Therefore 
\begin{equation}
F(\rho)\simeq c_0 \,e^{2\rho} (2\rho)^{-\xi/2-L-1}\, 
\end{equation}
for large $\rho$. This is the correct behavior of the series, that of course leads to an unacceptable wave function, unless the series 
has a finite number of  non-vanishing terms. 
We leave the details for the reader. She/he could also address the problem of a particle in a piecewise Coulomb potential given by
\begin{equation}
\label{ppotentialcoul}
V(r)=\left\{\begin{matrix}
-\frac{k}{R} \,\,\,\,\, &&0<r<R, \\ 
-\frac{k}{r} \,\, && r>R,
\end{matrix}\right.
\end{equation} 
following the procedure described for the piecewise harmonic oscillator.

\section{Conclusions}

We have discussed in detail the asymptotic behavior of the solutions of the Schroedinger equation with harmonic-like potentials. Following the standard approach we looked for solutions of the form given in Eq.~\eqref{prop}. We have shown that when the even  and odd series contain an infinite number of terms, they have, up to a constant,  the same divergent asymptotic behavior as $z\to +\infty$, contrary to previous claims in many textbooks. This is a necessary property, given that there should be a linear combination of the odd and even series that produce  a solution that is convergent for $z\to +\infty$,  for any value of ${\mathcal E}$.

For the usual harmonic oscillator, Eq.~\eqref{prop} should be the solution to the Schroedinger equation for all values of $z$. If we choose $a_*$ such that the wave function converges at $z\to +\infty$, then it will diverge at $z\to -\infty$ (and viceversa). Therefore, the physically acceptable solutions are those for which both series contain a finite number of terms,
and ${\mathcal E}= n+1/2$.   However, for piecewise potentials, we can consider independent linear combinations of the even and odd series  in the regions $z<-l$ and $z>l$, such that $\vert \psi(z)\vert^2$ is integrable. The continuity conditions of the wave function and its first derivative at $z=\pm\,  l$ fix the allowed energy eigenvalues. We illustrated the procedure by computing the eigenvalues of a potential with a ``bathtub'' shape.

The main mathematical result in our discussion is the large-$\omega$ behavior of the series 
\begin{equation}
S(\omega)=\sum_{n=1}^\infty \frac{n^{-r}\omega^n}{n!}\simeq \omega^{-r}e^{\omega}\, ,
\end{equation}
that can be derived as described above and in the Appendix. It can be even checked numerically by the students  using Mathematica or similar programs, by plotting
$S(\omega) \omega^{r}e^{-\omega}$ as a function of $\omega$,  for different values of $r$.
\section*{Acknowledgements} This research was supported by ANPCyT, CONICET, and UNCuyo.

\section*{Appendix : Asymptotic behavior of the series $S(\omega)$}

In this Appendix we provide an alternative and more rigorous  proof of Eq.~\eqref{serieSSapprox}, which is the main mathematical ingredient in our work. 
The derivation is somewhat cumbersome, but only uses elementary bounds for different series.

For simplicity we will assume $r>0$ (the case $r<0$ can be treated using similar arguments).
Let us consider
\begin{equation}\label{serieSS}
e^{-\omega}\omega^r S(\omega)=e^{-\omega}\sum_{n=1}^\infty \left(\frac{\omega}{n}\right)^{r}\frac{\omega^n}{n!}\, ,
\end{equation}
and introduce the notation 
\begin{equation}\label{serieT}
T(\omega,n_1,n_2)=e^{-\omega}\sum_{n=n_1}^{n_2} \left(\frac{\omega}{n}\right)^{r}\frac{\omega^n}{n!}\, .
\end{equation}
We would like to see that $T(\omega,1,\infty)\to 1$ as $\omega\to\infty$.

On one hand, given any $0<\lambda<1$, we split the series as
\begin{equation}\label{split}
T(\omega,1,\infty)=T(\omega,1,[\lambda \omega]) + 
T(\omega, [\lambda \omega]+1,\infty)\, ,
\end{equation}
where the brackets denote integer part. As $\omega/n \leq \omega$, the first term can be bounded by
\begin{equation}\label{primercotaT}
T(\omega,1,[\lambda \omega]) \leq e^{-\omega}\omega^r\sum_{n=1}^{[\lambda\omega]}\frac{\omega^n}{n!}.
\end{equation}
Noting that $\omega^n/n!$ is an increasing function of $n$ for $n\leq [\omega]$, we see that the series on the right hand side of Eq.~\eqref{primercotaT} satisfies
\begin{equation}\label{cotarhs}
\sum_{n=1}^{[\lambda\omega]}\frac{\omega^n}{n!} \leq \sum_{n=1}^{[\lambda\omega]}\frac{\omega^{[\lambda\omega]}}{[\lambda\omega]!} = [\lambda\omega]  \frac{\omega^{[\lambda\omega]}}{[\lambda\omega]!} \leq \lambda \omega \frac{\omega^{[\lambda\omega]}}{[\lambda\omega]!}
\end{equation}
and, hence, putting Eq.~\eqref{primercotaT} and Eq.\eqref{cotarhs} together we obtain $T(\omega,1,[\lambda \omega]) \leq  \lambda e^{-\omega} \omega^{r+1}
\omega^{[\lambda\omega]}/[\lambda\omega]!$. Let us show that $\lambda e^{-\omega} \omega^{r+1}
\omega^{[\lambda\omega]}/[\lambda\omega]!$ and, hence,  $T(\omega,1,[\lambda \omega])$, vanishes as $\omega\to\infty$. Using Stirling's approximation we see that, for large $\omega$,
\begin{equation}
e^{-\omega}\frac{\omega^{[\lambda\omega]}}{[\lambda\omega]!} \simeq e^{-\omega}\frac{\omega^{[\lambda\omega]}}{\sqrt{2\pi [\lambda\omega]}} \left(\frac{e}{[\lambda\omega]}\right)^{[\lambda\omega]}.
\end{equation}
Then, observing that 
\begin{equation}
\left(\frac{\omega}{[\lambda\omega]}\right)^{[\lambda\omega]} \lesssim \left(\frac{1}{\lambda}\right)^{\lambda\omega}
\end{equation}
and $\sqrt{2\pi [\lambda\omega]} \simeq \sqrt{2\pi \lambda\omega}$, we obtain
\begin{equation}\label{zeroexponentially}
e^{-\omega}\frac{\omega^{[\lambda\omega]}}{[\lambda\omega]!} \lesssim \frac{e^{-\omega}}{\sqrt{2\pi \lambda\omega}} \left(\frac{e}{\lambda}\right)^{\lambda\omega}
\end{equation}
and, since $\left(e/\lambda\right)^{\lambda} < e$ for $0<\lambda < 1$, we deduce from Eq.~\eqref{zeroexponentially} that $e^{-\omega}\omega^{[\lambda\omega]}/[\lambda\omega]!$ goes to zero exponentially as $\omega \to \infty$. This proves that $\lambda e^{-\omega} \omega^{r+1}
\omega^{[\lambda\omega]}/[\lambda\omega]!$ vanishes as $\omega \to \infty$. Therefore the first term in Eq.~\eqref{split} also vanishes.

The second term in Eq.~\eqref{split} can be bounded by
\begin{equation}\label{segundacotaT}
T(\omega, [\lambda \omega]+1,\infty)\leq e^{-\omega}\left(\frac{\omega}{[\lambda\omega]+1} \right)^r\sum_{n=[\lambda\omega]+1}^\infty\frac{\omega^n}{n!}
\end{equation}
simply noting that $\omega/n \leq \omega/([\lambda\omega]+1)$. Since $\omega/([\lambda\omega]+1) \leq 1/\lambda$ and 
\begin{equation}
\sum_{n=[\lambda\omega]+1}^\infty\frac{\omega^n}{n!} \leq e^{\omega},
\end{equation}
we deduce  that
$T(\omega, [\lambda \omega]+1,\infty) \leq \left(1/\lambda\right)^r$
and, therefore,
\begin{equation}\label{final1}
T(\omega,1,\infty)\leq T(\omega,1,[\lambda \omega]) + \left(\frac{1}{\lambda}\right)^r \xrightarrow[\omega\to\infty]{}\, \left(\frac{1}{\lambda}\right)^r.
\end{equation}
On the other hand, given any $\sigma>1$ we have 
\begin{equation}\label{bound3}
T(\omega,1,\infty)\geq T(\omega,1,[\sigma\omega])
\geq  e^{-\omega}\left(\frac{\omega}{[\sigma\omega]}\right)^r
\, \sum_{n=1}^{[\sigma\omega]}\frac{\omega^n}{n!}.
\end{equation}
We will see that $e^{-\omega}\, \sum_{n=1}^{[\sigma\omega]}\omega^n/n!$ tends to one as $\omega \to \infty$.
Given that
\begin{equation}\label{bound2}
e^{-\omega}\, \sum_{n=1}^{[\sigma\omega]}\frac{\omega^n}{n!}
=  e^{-\omega} \left(e^\omega-1-\sum_{[\sigma\omega]+1}^\infty\frac{\omega^n}{n!}\right) = 1 - e^{-\omega} - e^{-\omega}\sum_{[\sigma\omega]+1}^\infty\frac{\omega^n}{n!},
\end{equation}
it suffices to show that $e^{-\omega}\sum_{[\sigma\omega]+1}^\infty\omega^n/n!$ vanishes as $\omega \to \infty$. Now, since
\begin{eqnarray}
\sum_{[\sigma\omega]+1}^\infty\frac{\omega^n}{n!} &=& \frac{\omega^{[\sigma\omega]+1}}{([\sigma\omega]+1)!} \left(1+\frac{\omega}{[\sigma\omega]+2}+ \frac{\omega^2}{([\sigma\omega]+3)([\sigma\omega]+2)}+\cdots \right) \nonumber\\
&\leq& \frac{\omega^{[\sigma\omega]+1}}{([\sigma\omega]+1)!}  \left(1+\frac{\omega}{[\sigma\omega]+1}+ \frac{\omega^2}{([\sigma\omega]+1)^2}+\cdots \right)
\end{eqnarray}
and $\omega/([\sigma\omega]+1) \leq 1/\sigma$, we deduce
\begin{eqnarray}\label{bound4}
e^{-\omega}\sum_{[\sigma\omega]+1}^\infty\frac{\omega^n}{n!}
\leq e^{-\omega}\frac{\omega^{[\sigma\omega]+1}}{([\sigma\omega]+1)!}
\sum_{k\geq 0}\left(\frac{1}{_{\sigma}}\right)^k
= e^{-\omega}\frac{\sigma}{\sigma -1}
\frac{\omega^{[\sigma\omega]+1}}{([\sigma\omega]+1)!}\, .
\end{eqnarray}
Once more, one can check that this last term vanishes as $\omega\to\infty$. Then, the left hand side of Eq.~\eqref{bound2} tends to one as $\omega \to \infty$ and, therefore, from Eq.~\eqref{bound3} we see that
\begin{equation}
\label{final2}
T(\omega,1,\infty)\geq   e^{-\omega}\left(\frac{\omega}{[\sigma\omega]}\right)^r
\, \sum_{n=1}^{[\sigma\omega]}\frac{\omega^n}{n!} \xrightarrow[\omega\to\infty]{}\, \left(\frac{1}{\sigma}\right)^r.
\end{equation}
Now, since $\lambda$ and $\sigma$ were arbitrarily close to $1$ (from below and above, respectively),  Eq.~\eqref{final1} and Eq.~\eqref{final2} imply that
$
\lim_{\omega\to\infty} T(\omega,1,\infty) =1\, ,
$
which is the desired statement.

\end{document}